\documentclass{Interspeech2024}
\usepackage{multirow}

\interspeechcameraready

\title{VoxMed: One-Step Respiratory Disease Classifier using Digital Stethoscope Sounds }

\name[]{Paridhi}{Mundra*}
\name[]{Manik}{Sharma*}
\name[]{Yashwardhan}{Chaudhuri*}
\name[]{Orchid Chetia}{Phukan}
\name[]{Arun Balaji}{Buduru}

\address{
  IIIT-Delhi, India\\ *equal contribution
}
\email{\{paridhi20392, yashwardhan20417, manik21336, orchidp, arunb\}@iiitd.ac.in}

\keywords{speech recognition, Disease Detection, }

\begin{document}

\maketitle

\begin{abstract}
As respiratory illnesses become more common, it is crucial to quickly and accurately detect them to improve patient care. There is a need for improved diagnostic methods for immediate medical assessments for optimal patient outcomes. This paper introduces VoxMed, a UI-assisted one-step classifier that uses digital stethoscope recordings to diagnose respiratory diseases. It employs an Audio Spectrogram Transformer(AST) for feature extraction and a 1-D CNN-based architecture to classify respiratory diseases, offering professionals information regarding their patients respiratory health in seconds. We use the ICBHI dataset, which includes stethoscope recordings collected from patients in Greece and Portugal, to classify respiratory diseases. GitHub repository: https://github.com/Sample-User131001/VoxMed

\end{abstract}

\section{Introduction}
Respiratory-related diseases, such as chronic obstructive pulmonary disease and asthma, are leading causes of mortality globally \cite{cukic2012asthma}, making a swift and dependable respiratory diagnosis important for averting any respiratory complication.

We introduce VoxMed, a simple UI system that detects respiratory diseases using digital stethoscope sounds. Our goal is to overcome the challenges of traditional respiratory diagnostics, which typically involve long waits in crowded clinical settings\cite{brekke2008competition}. With VoxMed, healthcare practitioners can efficiently respond to patients' medical needs based on fast respiratory health assessments.
Past studies\cite{paraschiv2020machine} have shown that machine learning algorithms can potentially evaluate medical audio data to identify illnesses\cite{gairola2021respirenet}. VoxMed elaborates on the previous works to develop a one-step interface that enables doctors to record patient stethoscope sounds and support seamless diagnosis and analysis of patients. VoxMed accomplishes this by employing an Audio Spectrogram Transformer(AST)\cite{gong21b_interspeech} to extract features and a 1-D Convolutional Neural Network (CNN) architecture for classifying diseases, as shown in Figure 1. 
We test VoxMed's performance using the ICBHI challenge dataset\cite{rocha2018alpha}, featuring stethoscope sounds gathered from patients in Greece and Portugal. Our findings demonstrate that VoxMed reliably identifies respiratory diseases such as COPD, potentially enhancing patient care.
Our findings illustrate the platform's ability to reliably identify respiratory disorders, underlining its potential as a useful tool for respiratory disease diagnosis and therapy in clinical practice.
\begin{figure}[h]
    \centering
    \includegraphics[width=0.40\textwidth]{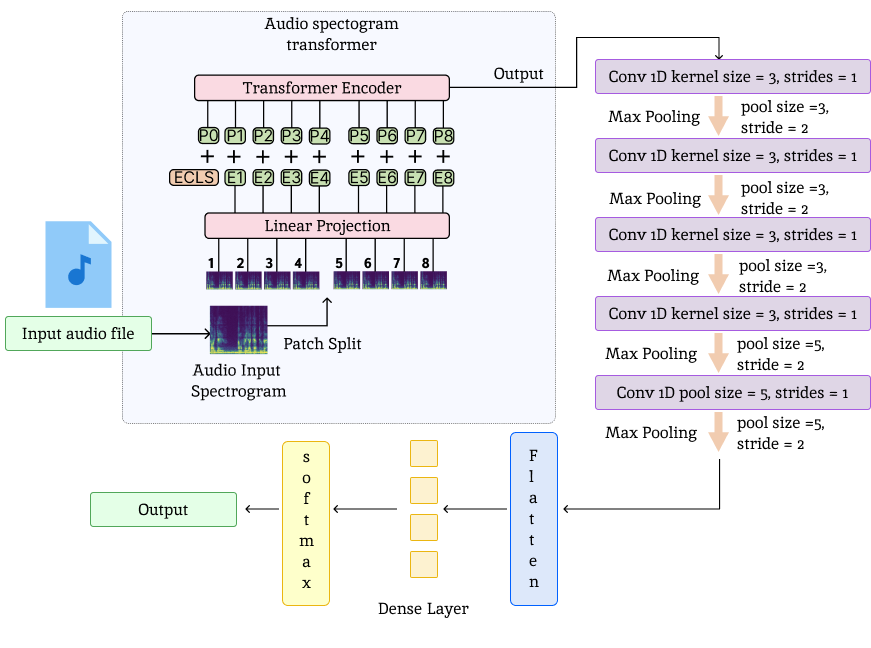}
    \caption{\textbf{VoxMed Architecture:} The architecture accepts a wav file for digital stethoscope sound and passes through an audio spectrogram transformer to extract features. Extracted features are passed through a 1-D CNN architecture as shown in the image to detect the type of respiratory disease.}
    \label{fig:example}
\end{figure}

\section{Application Overview}
Our work uses a simple 1-step respiratory disease detection process, illustrated in Figure 2, where users submit digital stethoscope sound samples of any duration. The model internally processes the audio using AST to make audio embeddings that are passed to a 1D-CNN, for identifying potential patient diseases. Our current application categorizes diseases as Healthy, COPD, or other potential diseases like URTI, LRTI, and Bronchitis.

\textbf{Data Details}:
We utilize the ICBHI Sound Database\cite{rocha2018alpha} for our implementation, comprising 920 annotated recordings from 126 patients, totalling around 6898 respiratory cycles. The database provides comments for crackles, wheezes, and combinations, offering clean and noisy respiratory recordings reflecting real-life scenarios and featuring patients of diverse ages.
\begin{figure*}[h]
    \centering
    \includegraphics[width=0.85\textwidth]{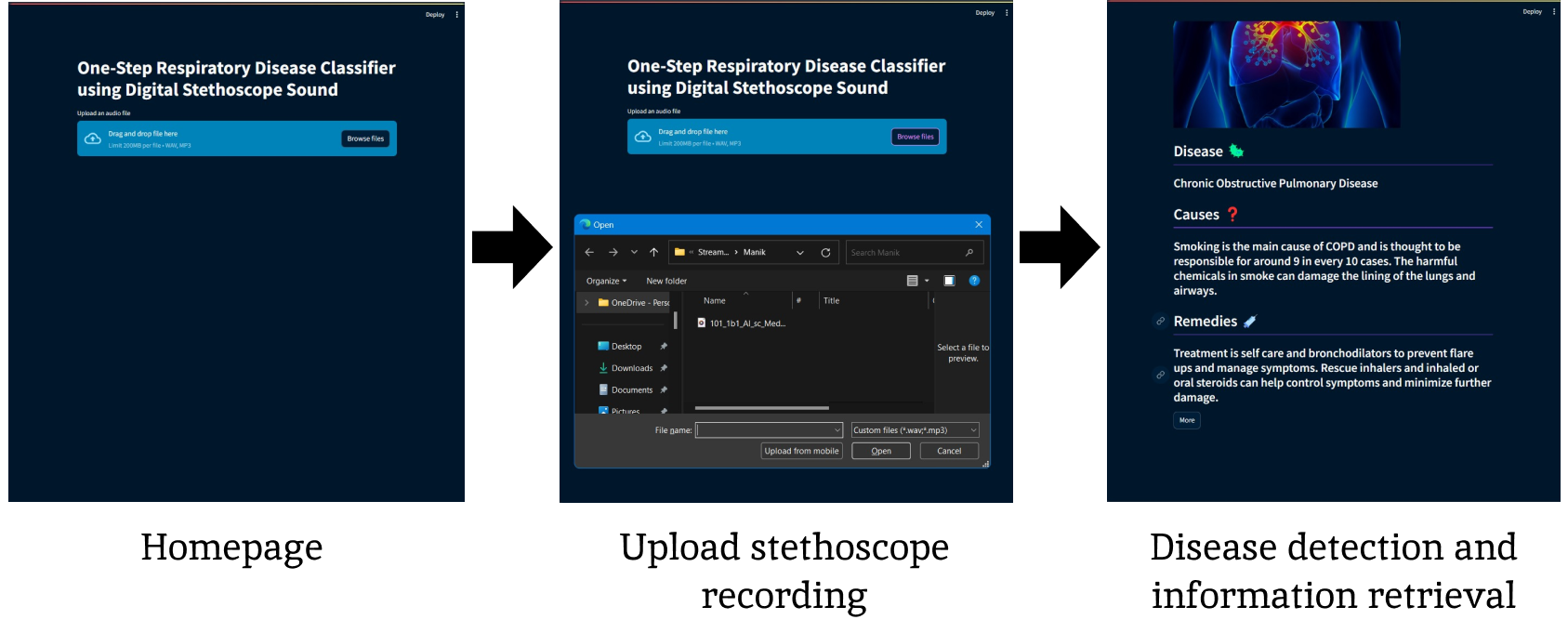}
    \caption{\textbf{VoxMed Workflow:} VoxMed respiratory detection requires the input of a digital stethoscope recording from the patient. We upload the recording through the UI and click on submit to process. Finally, we see possible respiratory ailment, its symptoms, and more information gathered through scraping information using APIs.}
    \label{fig:example}
\end{figure*}
\textbf{Feature Extraction and Model Architecture}:
Our model utilizes AST for audio embedding generation, operating at a 16000 sampling rate. Figure 1 showcases the CNN architecture tailored for classification tasks, featuring multiple layers like convolutional, max-pooling, and fully connected layers. The input layer employs a 1D convolutional layer with 256 filters and a size of 3, followed by ReLU activation. Subsequent layers include decreasing filter sizes and max-pooling, with dropout regularization applied to prevent overfitting. The final layer consists of a softmax activation function with six units corresponding to classification classes.

\textbf{Model Compilation and Training}: The model is built with the Adam optimizer and a categorical cross-entropy loss function. we use an 80-20 split to train and evaluate our model.

\section{Evaluation}
\textbf{Ablations:}
We assess VoxMed's performance on a class-by-class basis, utilizing macro weighted F1 score and accuracy metrics, as detailed in Table 1. Experimentation involved testing various feature extraction networks within VoxMed, including Wav2Vec2\cite{baevski2020wav2vec}, Unispeech\cite{wang2021unispeech}, AST\cite{gong21b_interspeech}, and WavLM\cite{chen2022wavlm}. Our analysis reveals that AST consistently outperforms the other methods across both Macro F1 and accuracy measurements. Notably, the classification criterion of Healthy/COPD/others demonstrates the highest F1 score, suggesting heightened reliability in disease detection using this specific approach.
\begin{table}[]
\centering
\begin{tabular}{lll}
\hline

\textcolor{blue}{\textbf{Models}}                                                & 
\textcolor{blue}{\textbf{Accuracy}} & 
\textcolor{blue}{\textbf{F1}}   \\ \hline \\
\multicolumn{3}{l}{\textbf{Healthy/COPD/others}}                                   \\ \hline
\textbf{AST(VoxMed)}\cite{gong21b_interspeech} & \textbf{0.90}     & \textbf{0.70} \\ \hline
Wav2Vec2\cite{baevski2020wav2vec}      & 0.88     & 0.48 \\ \hline
Unispeech\cite{wang2021unispeech}      & 0.89     & 0.58 \\ \hline
WavLM\cite{chen2022wavlm}              & 0.86     & 0.47 \\ \hline \\
\multicolumn{3}{l}{\textbf{Healthy/COPD/URTI/others}}                              \\ \hline
\textbf{AST(VoxMed)}\cite{gong21b_interspeech} & \textbf{0.85}     & \textbf{0.69} \\ \hline
Wav2Vec2\cite{baevski2020wav2vec}      & 0.90     & 0.67 \\ \hline
Unispeech\cite{wang2021unispeech}      & 0.86     & 0.43 \\ \hline
WavLM\cite{chen2022wavlm}              & 0.86     & 0.44 \\ \hline \\
\multicolumn{3}{l}{\textbf{Healthy/COPD/URTI/LRTI/Bronchitis}}                              \\ \hline
\textbf{AST(VoxMed)}\cite{gong21b_interspeech} & \textbf{0.90}     & \textbf{0.67} \\ \hline
Wav2Vec2\cite{baevski2020wav2vec}      & 0.90     & 0.48 \\ \hline
Unispeech\cite{wang2021unispeech}      & 0.89     & 0.54 \\ \hline
WavLM\cite{chen2022wavlm}              & 0.89     & 0.53 \\ \hline
\end{tabular}
\caption{Ablations: Class-wise performance of VoxMed With different feature extraction backbones; F1 is macro-average F1 score.}
\end{table}

\section{Conclusion}

VoxMed emerges as a solution for rapid respiratory disease diagnosis, offering healthcare professionals immediate insights into patient conditions. By leveraging an AST feature extraction and a 1-D CNN-based classifier, VoxMed achieves competitive performance along with a seamless 1-step UI. Its ability to streamline diagnostic processes and provide timely medical assessments underscores its potential to significantly impact patient care and healthcare delivery.

\bibliographystyle{IEEEtran}
\bibliography{main}

\end{document}